\documentclass[12pt]{article}
\usepackage[english]{babel}
\usepackage[utf8]{inputenc}
\usepackage{multicol}
\usepackage{amsfonts,amsmath,amssymb}
\usepackage{graphics}
\usepackage{graphicx}

\title{Effects of density-dependent migration on a population subjected to Allee effect}
\author{G.A. Ossandón \& Ricardo Castro Santis\\
\small{Departamento de Matemática}\\
\small{Universidad Tecnológica Metropolitana, Santiago, Chile}\\
\small{gusosar@utem.cl , rcastro@utem.cl}}
\date{}
\begin{document}
\maketitle

\begin{abstract}
This study assessed the effects of migration on the dynamics of a species population. It was considered that the species in its natural state and without the presence of migration exhibited Allee effect. This work also considered migration as a density-dependent function, which, from a maximum rate, decreases to a minimum of zero when the population reaches its carrying capacity.

\end{abstract}

\section{Introducción}\label{section}

A recurring problem in situations of endangered species preservation is that mere protection through hunting prohibition is not sufficient to ensure the recovery of the population. This fact occurs when population density decreases to a level in which reproduction and maintenance becomes very difficult, thus decreasing the reproductive rate to negative levels, which will finally end in the extinction of the species. This phenomenon is called \textit{Efecto Allee} \cite{ALLE}\cite{E0}\cite{E1}.In such a situation, it is necessary to intervene in order to promote the recovery of the population levels that will allow the species to survive in a natural way. One of these interventions is the introduction of new individuals in the habitat of the species. This strategy is often effective; however, it has high costs. This way, it is necessary to perform it in a way able to optimise the outcomes and resources\cite{BC,CKI}. Within this context, a natural way is to generate a migratory flow dependent on the population level of the species, so that migration is greater than low population levels and ceases completely when the population reaches the carrying capacity of the environment. There are many studies on models with Alle effect, including deterministic and stochastic models, in ordinary differential equations and in partial derivatives. It is important to point out that the use of differential equations in partial derivatives is one of the possible approaches, very useful when the Allee effect is produced by the low rate of encounters between individuals. In Petroski\&Lian\cite{petrovskii2003exactly} they propose a model with unidimensional spatial diffusion with Allee effect and dense dependent migration, whose migration function is a linear function. This considered migration is the product of environmental factors and biological mechanisms.\\

The first thing to observe, and perhaps most importantly, is that it is not possible to maintain a migration above the carrying capacity of the environment, as the habitat does not support a larger population , will cause the surplus to be quickly eliminated. From this point of view it is natural that a migratory flow ceases when the population reaches the carrying capacity.\\
  
These hypotheses can be translated mathematically in the following way: Let $f(x)$ be the immigration rate. We will consider $f$ as a decreasing function in the variable $x$ such that $f(K) = 0$, where $K$ is the capacity of environment. Therefore
    
   $$\displaystyle\max_{x\in[0,K]}\{f(x)\}=f(0),$$
   
If we call $f(0)=\alpha$, the migration rate is a decreasing function joining the points $(0, \alpha)$ with the point $(K, 0)$\\

In other words, if $K$ indicates the carrying capacity of the environment inhabited by the species, $x(t)$ will refer to the population of the species in time $t>0$, considering a positive and continuous function of migration rate $f_{\alpha}(x)$, strictly decreasing in the interval $[0,K]$, so that $f_{\alpha}(0)=\alpha$, and $f_{\alpha}(K)=0$. A natural choice in this context is to assume that the $f_{\alpha}$ function is linear, as shown in Figure 1.
\begin{center}
 \includegraphics[width=8cm]{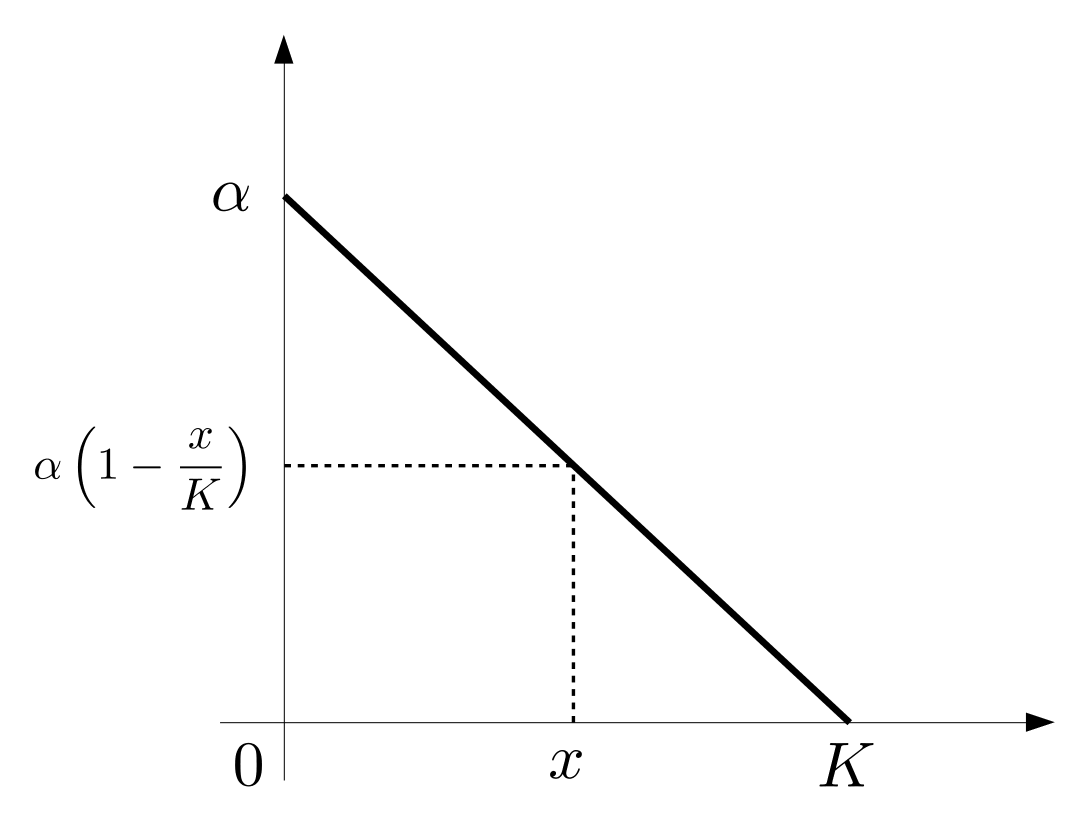}\\
Figure 1: Inmigration rate.
\end{center}

Therefore, the explicit equation of $f_{\alpha}$ is:
 \begin{equation}\label{eq:inmigracion}
f_\alpha(x)=\alpha\left(1-\frac{x}{K}\right)\,\, , \,\, \forall \,\, x\in [0,K]
\end{equation}

The parameter $\alpha$ represents the maximum value of the migration rate, which occurs in the total absence
of population in the environment, which can be interpreted as the colonization rate of the habitat. The
carrying capacity $K$ will be considered as a fixed parameter exclusively dependent on environmental
conditions and that is not affected by the population of the species.

In the present study, we considered a logistic type of natural growth of the species subjected to Allee
effect \cite{BC}\cite{CKI}\cite{EK} at a population level $m$, with intrinsic rate growth $r$, and carrying capacity $K$, i.e., the growth
rate of the species in the absence of migration will be:

\begin{equation}\label{eq:Allee}
 g(x)=r\left(1-\frac{x}{K}\right)(x-m)x\quad\mbox{with}\quad 0<m<K\quad \mbox{and}\quad r>0
\end{equation}

\section{The Model}

Taken into consideration the hypotheses discused in the introduction, the model for population
dynamics will have the following equation:
\begin{equation}\label{eq:model}
\left\{\begin{array}{rcl}
 \dfrac{d}{dt}x(t)&=& g(x)+f_\alpha(x)\\
 x(0)&=&x_0
 \end{array}\right.
\end{equation}
where the initial population $x_{0} \in [0,K]$, and the functions $g(x)$ and $f_{\alpha}(x)$ are those defined in equations
\eqref{eq:Allee} and \eqref{eq:inmigracion}, respectively.\\

Equation \eqref{eq:model} can be written as follows:
\[\begin{array}{rcl}
   x'(t)&=&g(x)+f_\alpha(x)\\
   {}\\
   {}&=&r\left(1-\dfrac{x}{K}\right)(x-m)x+\alpha\left(1-\dfrac{x}{K}\right)\\
   {}\\
   {}&=&\left(1-\dfrac{x}{K}\right)\left(r(x-m)x+\alpha\right)\\
   {}\\
   {}&=&\left(1-\dfrac{x}{K}\right)\left(rx^2-mrx+\alpha\right)
  \end{array}
\]
which indicates that $x=K$ remains an equilibrium point of the system. The other equilibrium points
will depend on the quadratic factor $(rx^{2}-mrx+\alpha)$.\\

A discriminant analysis of equation  $rx^{2}-mrx+\alpha$ leads to:
\begin{equation}\label{eq:discriminante}
 \Delta=m^2r^2-4r\alpha
\end{equation}

This equation divides the analysis into three situations relating to the possible values that can achieve
maximum migration rate $\alpha$ with respect to the intrinsic growth rate $r$ and the Allee level $m$. This fact is
illustrated in the graph of the phase diagram, as shown in the figures 2, 3 nd 4:
\\

\begin{center}
 \includegraphics[width=3cm]{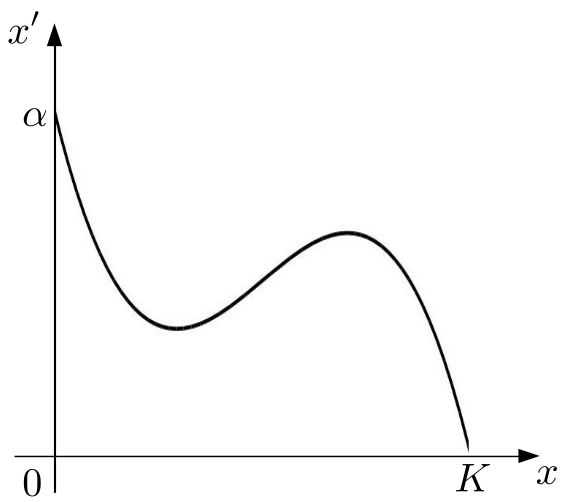}\hspace{0.5cm}
 \includegraphics[width=3cm]{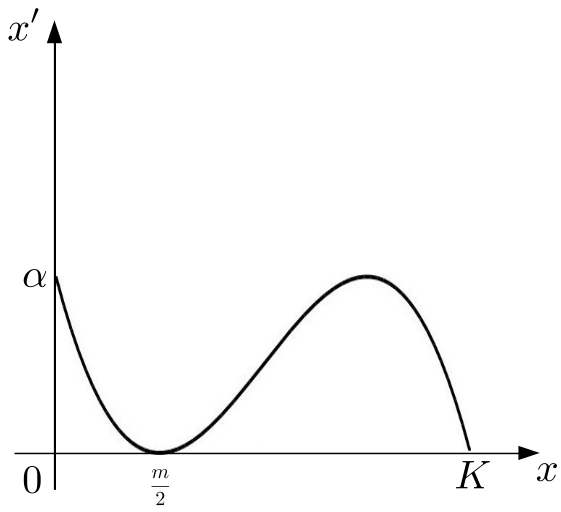}\hspace{0.5cm}
 \includegraphics[width=3cm]{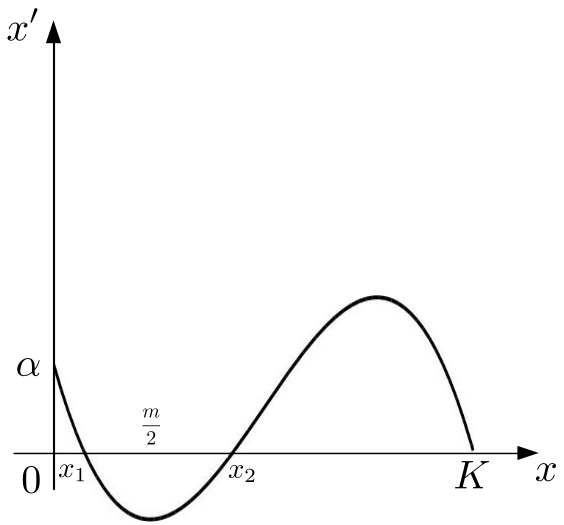}\\
Figure 2\hspace{2.5cm}Figure 3\hspace{2.5cm}Figure 4\\
$\Delta>0$\hspace{2.8cm}$\Delta=0$\hspace{2.8cm}$\Delta<0$
\end{center}

\noindent\textbf{NOTE:}In this model, $x=0$ is not the equilibrium point of the system. This way, it is possible that the
initial population rate of the species is zero. In this situation, it can be affirmed that there is a
colonization of the habitat by the migratory flow.

\subsection{Case $\alpha>\dfrac{m^2r}{4}$}

If $\alpha>\frac{m^{2}r}{4}$, the discriminant factor defined in equation \eqref{eq:discriminante} will be negative and the only equilibrium point
of the system will be $x=K$, as illustrated in Figure 2. In this condition $x'(t)>0$ for all $x_0 \in [0,K]$,
therefore, the population will grow from the initial value $x_{0}$ to the carrying capacity $K$.\\

In this situation, for any initial value, including zero initial population, the species will asymptotically
tend to the carrying capacity of the environment.\\

Figure 5 shows the temporal evolution of the population of the species, which starts with a $x_{0}$ value.\\

\begin{minipage}{12cm}
\begin{center}
 \includegraphics[width=6cm]{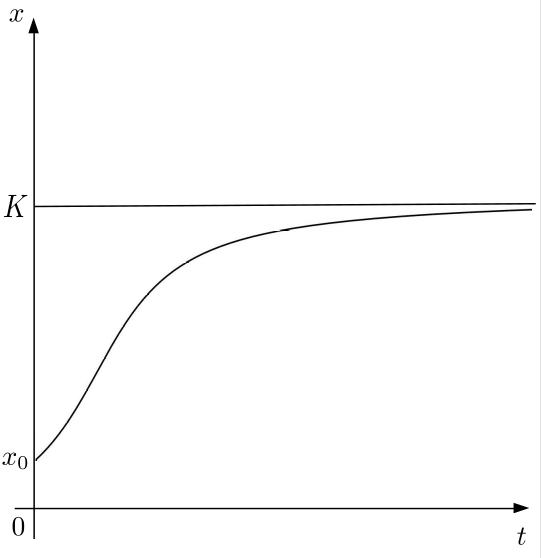}\\
Figure 5: Trajectory case $\Delta<0$
\end{center}
\end{minipage}

\subsection{Case $\alpha=\dfrac{m^2r}{4}$}
There is a new equilibrium point as a solution to equation $rx^{2}-mrx+\alpha=0$, which, with the
condition $\alpha=\frac{m^2r}{4}$, becomes $\,r\left(x-\frac{m}{2}\right)^2=0$, thus obtaining the equilibrium point $x=\frac{m}{2}$, which corresponds
to half the value of the natural Allee level of the species. The equilibrium point  $x=\frac{m}{2}$ is unstable of
attractor-repulsor design.

As shown in Figure 3, for all $x_0\in[0,K]$; the trajectories will be non-decreasing for any
initial value $0\le x_0<K$.\\

If $0\le x_0<\frac{m}{2}$, the trajectories will grow asymptotically at $\frac{m}{2}$, whereas, if $\frac{m}{2}<x_0<K$, the
trajectories will grow asymptotically with respect to carrying capacity $K$.\\

In ecological terms, the species will finally achieve its recovery. Even though the initial population is
less than $\frac{m}{2}$, a positive disturbance may eventually raise that level. This way, final recovery will start to
achieve carrying capacity, as shown in Figure 6.
\begin{center}
 \includegraphics[width=6cm]{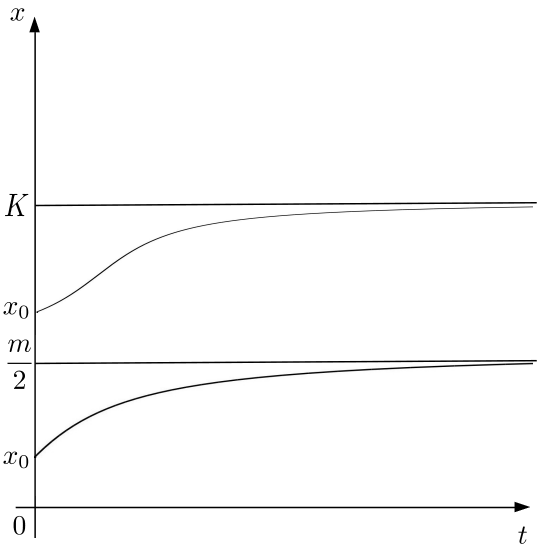}\\
Figure 6: : Trajectory case $\Delta=0$
\end{center}

\subsection{Case $\alpha<\dfrac{m^2r}{4}$}

This case exhibits the greatest dynamic richness. There are two equilibrium points relating to value $\frac{m}{2}$. The discriminant $\Delta=m^2r^2-4r\alpha>0$ and the equation solutions $rx^2-mrx+\alpha=0$ will be

\begin{equation}\label{eq:equi-point}
 x_1=\frac{m}{2}-\frac{\sqrt{\Delta}}{2r}\qquad\mbox{y}\qquad x_2=\frac{m}{2}+\frac{\sqrt{\Delta}}{2r}
\end{equation}
as  $\displaystyle\sqrt{m^2r^2-4\alpha}<mr$, the following inequalities are obtained

$$0<x_1<\dfrac{m}{2}<x_2<m$$

Figure 4 illustrates the situation described, from which it can be easily observed that $x'>0$\quad if\quad $0\le x<x_1$\quad or\quad $x_2<x<K$ and, conversely, $x'<0$\quad if\quad $x_1<x<x_2$. Therefore, the trajectories will
have the following behaviour depending on the initial population $x_0$ value.

\begin{itemize}
 \item If $0\le x_0<x_1$, the population will asymptotically grow to level $x_1$.
 \item If $x_1<x_0<x_2$, the population will asymptotically decrease to level $x_1$.
 \item If $x_2<x_0<K$, the population will asymptotically grow to carrying capacity $K$.
\end{itemize}

Figure 7 illustrates the situation described.
\begin{center}
 \includegraphics[width=6cm]{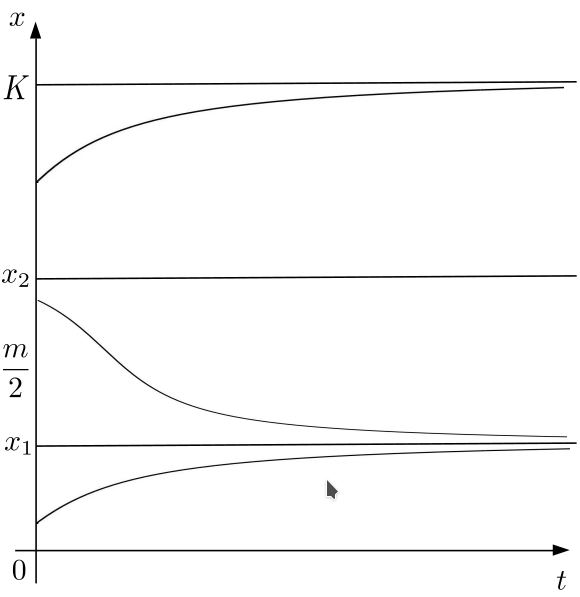}\\
Figure 7: : Trajectory case $\Delta>0$
\end{center}

\section{Some conclusions}

\begin{itemize}
 \item The Allee level, which, in the absence of migration is related to a population abundance of $m$,
decreases to the value $x_2 \in\left]\frac{m}{2},m\right[$.

 \item The presence of migration prevents the population from becoming extinct, with an attractor
equilibrium point of $x_1\in\left]0,\frac{m}{2}\right[$.

 \item This equilibrium is exclusively maintained by the migratory flow and can be interpreted as
precarious subsistence equilibrium, because, since the migrant population is far below the
natural Allee effect level, it fails to survive and is continuously replaced by new migrants,
which will never be able to recover the population with respect to the carrying capacity of the
environment.

 \item From the perspective of the problem of induced immigration as a technique for recovery of
species, it can be observed that, if the initial population is very small $\frac{m}{2}$, the only
possibility for recovering will be a strong migratory flow, i.e., a rate given by $\alpha>\frac{m^2r}{4}$.

 \item If the initial population is greater than\quad $\frac{m}{2}$\quad but less than  $m$ , a migration rate will be necessary to
ensure that\quad $x_0>x_2$\quad and the following result will be obtained from equation \eqref{eq:equi-point}: $\alpha> rx_{0}(m-x_{0})$

\end{itemize}

\end{document}